\documentclass[a4paper,fleqn,usenatbib, referee]{mnras}
\usepackage{ulem}
\newcommand{\II}{\small{II}\normalsize}
\usepackage{subfigure}

\usepackage[T1]{fontenc}
\usepackage{ae,aecompl}

\usepackage{graphicx}	
\usepackage{amsmath}	
\usepackage{amssymb}	
\usepackage{subfigure}
\usepackage{natbib}
\usepackage{adjustbox}
\usepackage{helvet,soul}

\title[A possible cycle for AU Mic]{First long-term activity study of AU Microscopii: a possible chromospheric cycle.}

\author[Iba\~nez Bustos et al.]{R. V. Iba\~nez Bustos$^{1}$\thanks{Contact e-mail: ribanez@iafe.uba.ar}\thanks{Based on data obtained at Complejo Astron\'omico El Leoncito, operated under agreement between the Consejo Nacional de Investigaciones Cient\'\i ficas y T\'ecnicas de la Rep\'ublica Argentina and the National Universities of La Plata, C\'ordoba and San Juan.}, A. P. Buccino$^{1,2}$, M. Flores$^{3, 4}$, C. I. Martinez$^{3}$ 
\newauthor D. Maizel$^{1}$, Sergio Messina $^{5}$ and P. J. D. Mauas$^{1,2}$ 
\\
$^{1}$Instituto de Astronom\'\i a y F\'\i sica del Espacio (CONICET-UBA), C.C. 67 Sucursal 28, C1428EHA-Buenos Aires, Argentina. \\
$^{2}$Departamento de F\'\i sica. FCEyN-Universidad de Buenos Aires, Buenos, Argentina.\\
$^{3}$Instituto de Ciencias Astron\'omicas, de la Tierra y del Espacio (ICATE-CONICET), San Juan, Argentina.\\
$^{4}$Facultad de Ciencias Exactas, F\'isicas y Naturales, Universidad Nacional de San Juan, San Juan, Argentina.\\
$^{5}$ INAF-Catania Astrophysical Observatory, via S. Sofia 78, 95123, Catania, Italy
}

\date{Last updated 2018 May 2th}

\pubyear{2018}

\begin{document}
\label{firstpage}
\pagerange{\pageref{firstpage}--\pageref{lastpage}}
\maketitle

\begin{abstract}
M stars are ideal targets to search for Earth-like planets. However, they usually have high levels of magnetic activity, which could affect their habitability and  make difficult the detection of exoplanets  orbiting around them. 
Unfortunately, long-term variability of dM stars has not been extensively studied, due to their low intrinsic brightness. For this reason, in 1999 we started the HK$\alpha$ project, which systematically observes the spectra of a large number of stars, in particular dM stars, at the Complejo Astron\'omico El Leoncito (CASLEO).  
In this work, we study the long-term activity of the young active dM1 star AU Microscopii. We analyze the Mount Wilson index $S$ derived from CASLEO spectra obtained between 2004 and 2016, which we complement with the $S$-index derived from HARPS, FEROS and UVES public spectra. 
We also analyze the simultaneous photometric counterpart provided by the ASAS public database for this star between 2000 and 2009, and our own photometry. 
In both totally independent time series, we detect a possible activity cycle of period $\sim$ 5 years. We also derived a precise rotation period for this star $P_{rot}= 4.85$ days, consistent with the literature. This activity cycle reflects that an $\alpha\Omega$ dynamo could be operating in this star.

\end{abstract}


\section{Introduction}
 
M stars, which represent 70\% of the stars in our Galaxy, have become
relevant targets  for different exo-planetary surveys during the last
decade (e.g. \citealt{Mayor09,Zechmeister09,Amado13,France16}). These
projects were mainly motivated by the fact that, due to their low
mass, M stars are favorable targets to detect Earth-like planets
around them. Furthermore, such planets seem to be frequent in the
solar neighborhood. By exploring the \textit{Kepler} transiting
exoplanets database \citep{Borucki10}, \cite{DressingCharbonneau15}
estimated a high occurrence rate of Earth and Super Earth-like planets
around M stars (2.5$\pm$0.2 planets per M star with radius between 1
and 4 Earth radii and period less than 200 days). Similar results were
also obtained from Radial Velocities (RV) surveys \citep{Bonfils13}. 

Another  motivation to study red dwarfs is that several active M stars are remarkably variable due to flare events (e.g. \citealt{Moffet74,Hilton11,Kowalski13}).   In these stars, a flare represents an  appreciable variation in luminosity with higher contrast than in solar-type stars. The released energy  during a  flare event could be eventually higher than  10$^{34}$ erg \citep{Hawley91, Mauas96}. A well-known tendency is that energetic flares are   less frequent  than small flares (e.g. \citealt{Lacy76,Hilton11}). The Kepler mission, designed to obtain high-precision and long-baseline light curves of numerous stars, brought further statistics to this field. \cite{Hawley14} found a strong correlation between flare energy, amplitude, duration, and decay time, with only a weak dependence on rise time. Furthermore, they  found that flares from inactive stars, though relatively rare, can be very energetic. Therefore, a terrestrial planet around an M  star could be continuously affected by small short flares and eventually by high-energy long flares constraining its habitability \citep{Buccino07,Vida17}. On the other hand, small flares from M dwarfs could play an important role in the atmospheric chemistry of orbiting planets (e.g. \citealt{Miguel15}).

Recently, several authors (e.g. \citealt{Shibata13,Kitchatinov16,Chang17,Yang17}) have explored  the role of  flare activity as a stellar activity indicator and thus its connection with the stellar dynamo, responsible for the generation and amplification of  stellar magnetic fields. The dynamo is thought to be driven by differential rotation ($\Omega$-effect) and turbulent convection ($\alpha$-effect) in the stellar interior \citep{Parker55,Parker63,Steenbeck69}. Several authors suggest that the tachocline - a narrow layer between the radiative core, which rotates rigidly, and the convection zone - is substantial in the solar dynamo to strengthen the magnetic field and to avoid magnetic field dissipation  (e.g. \citealt{Spiegel92}). 


In particular, it is believed that an $\alpha\Omega$-dynamo could be operating in the Sun and other late-F to mid-M stars \citep{Durney82} and this model could successfully reproduce the global magnetic activity in solar-type stars (e.g. \citealt{Sraibman17}). On the other hand, M dwarf stars with masses lower than 0.35M$_\odot$ (i.e. later than spectral type M4) are believed to be fully convective \citep{Chabrier97}. Therefore, their dynamo mechanism is expected to be different from a solar-type dynamo, as they do not have a tachocline (\citealt{Chabrier06,Browning08,Reiners10}).  However, \cite{Wright16} studied the X-ray activity-rotation relationship, which is a well-established proxy for the behaviour of the magnetic dynamo, for a set of purely convective low-rotators stars. They observed that the  X-ray emission from these stars correlates with their rotation periods in the same way as in Sun-like stars. One of the conclusions of that work is that the tachocline might not be a critical ingredient in the solar dynamo. Nevertheless, from spectroscopic  and \textbf{spectropolarimetric} observations, several authors infer that the magnetic topologies could change at the full convection boundary \citep{Donati08,Morin08,Reiners09}. Therefore, M stars are also interesting targets for the dynamo theory, as they bring the opportunity to explore if a solar-type dynamo could still operate beyond the convective threshold and also explore the magnetic topologies resulting from dynamo models.

Although M dwarfs have been widely studied for their flare activity during decades, their long-term activity is rather unexplored. For this reason, in 1999 we started the \textit{HK$\alpha$ Project} at the Complejo Astron\'omico El Leoncito (CASLEO), mainly dedicated to study the long-term chromospheric activity in bright partially and fully convective M stars. One of the main objectives of this project is to explore the stellar dynamo beyond the convective limit. From our data, we detected evidences of cyclic chromospheric activity in the early-M stars GJ 229 A and GJ 752 A \citep{Buccino11}, in three partially convective stars near the convective threshold; the binary system GJ 375 \citep{Diaz07} and GJ 388-AD Leo \citep{Buccino14} and the fully-convective active star Proxima Centauri \citep{Cincunegui07}. The exoplanetary survey \textit{High Accuracy Radial velocity Planet Searcher (HARPS)} expanded the detection of activity cycles in low-mass stars (e.g. \citealt{GomesdaSilva12,Diaz16}) in the southern neighborhood. By using spectroscopic and photometric observations, other programs also contributed to the detection of magnetic cycles in dM stars \citep{Robertson13,SuarezMascareno16,Wargelin17}. Furthermore,  theoretical and numerical studies also reproduce magnetic cycles in M stars (eg. \citealt{Kitchatinov14,Yadav16,Kuker18}). However, the statistics is still low in comparison to solar-type stars \citep{Baliunas95,Hall07,Lovis11,Egeland17,Flores17}. Therefore, the detection of new accurate activity cycles for low-mass stars could bring valuable observational evidence to explore the dynamo theory in this spectral class.

One of the targets extensively observed during the HK$\alpha$ Project is the M1Ve star AU Microscopii (GJ 803). AU Mic is a member of the $\beta$ Pictoris association, as inferred from space and Galactic velocity components, Lithium abundance and rotation period \citep{Messina17}. The association has a quoted age of 25$\pm$3\,Myr, which was recently derived by \cite{Messina16} with the lithium depletion boundary (LDB) method, and found to be in agreement with respect to the estimates from earlier works (e.g. \citealt{Mamajek14,Bell15}). It is a fast-rotator star with a $\sim$ 4.847 day-period \citep{Hebb07} and high-energy flares \citep{Cully93}. AU Mic is surrounded by an edge-on circumstellar debris disk that extends from 50 AU to 210 AU \citep{Kalas04,Liu04,Krist05}. 



Since 2004, we have been continuously observing this star as part of the ongoing  HK$\alpha$ project. This data-set allows us to have a  unique long time-series of Mount Wilson indexes, which is the most extended activity indicator used to detect stellar activity cycles \citep{Baliunas95,Metcalfe13}. The main purpose of the present work is to study for the first time the long-term chromospheric activity of AU Mic. In particular, in Section $\S$2 we present an overview of our spectroscopic observations, which are complemented by public spectroscopic data obtained form the European Southern Observatory (ESO) as well as from public photometry. We also obtained our own photometry with the Magnetic Activity and Transiting Exoplanets (MATE) telescope to acquire a precise rotation period for AU Mic.  In Section $\S$3 we describe our results, and finally we outline our discussion in Section $\S$4.

\section{Observations}\label{sec.obs}

The HK$\alpha$ Project is mainly dedicated to  systematically obtain mid-resolution echelle spectra ($R= \lambda / \delta\lambda \approx 13,000$) of a set of 150 dF5 to dM5.5 stars. The observations are obtained with the REOSC\footnote{http://www.casleo.gov.ar/instrumental/js-reosc.php} spectrograph, which is  mounted on the 2.15 m Jorge Sahade telescope at the Complejo Astron\'omico El Leoncito Observatory (CASLEO) in the Argentinian Andes. These echelle spectra were optimally extracted and flux calibrated using IRAF\footnote{The Image Reduction and Analysis Facility (IRAF) is distributed by the Association of Universities for Research in Astronomy (AURA),  Inc., under contract to the National Science Foundation} routines (see \citealt{Cincu04}, for details). At present, our database contains near 6000 mid-resolution spectra covering a  wavelength range between 3800 to 6900 \AA.

\begin{table}
\centering
\caption{Logs of the CASLEO Observations of AU Mic. Column 1 and 3:
  date of each observing run (MMYY). Column 2 and 4: Julian date xJD =
  JD - 2450000.} 
\begin{tabular}{cccccc}
\hline\hline\noalign{\smallskip}
Label	&	xJD &	&	& Label	&	xJD\\
\hline\noalign{\smallskip}
0904 & 3275 &	&	& 1009 & 5107 \\
0605 & 3543 &	&	& 0610 & 5357 \\
0805 & 3599 &	&	& 0910 & 5456 \\
1105 & 3698 &	&	& 0611 & 5730 \\
0906 & 3990 &	&	& 0911 & 5820 \\ 
0607 & 4281 &	&	& 0612 & 6089 \\
0907 & 4369 &	&	& 0912 & 6171 \\
0608 & 4638 &	&	& 0614 & 6833 \\
0908 & 4732 &	&	& 0914 & 6905 \\
0609 & 4990 &	&	& 0816 & 7614 \\

\hline 
\end{tabular}
\label{tabla_observaciones}
\end{table}

In Table \ref{tabla_observaciones} we show the observation logs of AU Mic  from CASLEO. The first and third column shows the date (month and year) of the observation and the second column lists $xJD = JD - 2450000$, where JD is the Julian date at the beginning of the observation. There is  a total of 20 individual observations distributed over twelve years between 2004 and 2016. 
Specific details of the observations are described in \cite{Cincu04}.

We complement our data with public high-resolution spectra ($R \sim 115.000$) obtained with the HARPS spectrograph attached to the 3.6 m ESO telescope. These observations, taken during two time intervals, $2003-2005$ and $2013-2016$, correspond to the ESO programs 072.C-0488(E), 075.C-0202(A) and 192.C-0224(A,B,C,H). These spectra have been automatically processed by the HARPS pipeline\footnote{http://www.eso.org/sci/facilities/lasilla/instruments/harps/}. 

In addition, we also included the Mount Wilson indexes  obtained with the FEROS spectrograph on the 2.2 m ESO telescope and the UVES spectrograph mounted at the Unit Telescope 2 (UT2) of the Very Large Telescope (VLT). The FEROS high-resolution spectra ($R \sim 48.000$) were obtained during the programs 072.A-9006(A), 075.A-9010(A), 178.D-0361(D) and 077.D-0247(A), already processed by the FEROS pipeline\footnote{http://www.eso.org/sci/facilities/lasilla/instruments/feros/}. We also incorporated the UVES high-resolution spectra ($R \sim 80.000$) of the programs 088.C-0506(A), 082.C-0218(A) and 075.C-0321(A) all reduced with the UVES procedure\footnote{http://www.eso.org/sci/facilities/paranal/instruments/uves/}. 


To perform a long-term analysis with an independent database, we also employed photometric observations provided by the All Sky Automated Survey (ASAS, \citealt{Pojmanski02}).  AU Mic was extensively observed between 2000 and 2009 by the ASAS-3 program. Similarly to our previous works \citep{Diaz07, Buccino11, Buccino14}, we choose the mean $V$ magnitude of each observing season as an activity proxy. We retained only the best quality data, and we discarded  all  observations  that  were  not qualified as either A or B in the ASAS database. 

Understanding the relationship between rotation and activity is
important in the study and interpretation of stellar dynamos. In order
to study the rotation and flare activity in AU Mic, we carried out a
photometric study of AU Mic during 2017, using a MEADE Ritchey-Chretien 16" telescope installed in the Carlos U. Cesco Height Station at the Observatorio Felix Aguilar (OAFA) in the Argentinian Andes in San Juan. This telescope, called \textit{MATE (Magnetic Activity and Transiting Exoplanets)}, is currently equipped with an SBIG STL11000M camera and Johnson BVRI filters. The effective CCD area is of 4008 $\times$ 2672  9$\mu$m pixels \citep{Schwartz13}. The CCD scale is 0.47 arcsec per pixel and, thus, the resulting Field of View is 31 arcmin $\times$ 21 arcmin. 

Since AU Mic is a bright star ($V=8.627$), it is hard to find suitable comparison stars in the Field of View (FoV). In this work, thanks to the wide field of the camera installed in MATE, we were able to simultaneously observe AU Mic and the A9V star HD 197673 of similar brightness ($V=8.52$),  centering the field at RA=20:45:44 DEC=-31:35:40 (J2000). We estimated the formal error for each observation as the sum in quadrature of the errors of the fluxes of the target and the reference star. After discarding saturated images and those with low signal-to-noise, we obtained 1902 images between February and November 2017, over non- consecutive nights to have a near uniform sampling throughout the expected rotation phase.  During each observing night, we obtained science images every five minutes.

We also obtained  5 to 8 \textit{sky flats} at the sunset and 20 \textit{bias} every night. We did not need to correct images by subtracting darks as their contribution was negligible after bias subtraction. We perform aperture photometry with the package \textsf{killastro} developed by our group. This package is based on the IRAF tasks  \textsf{ccdproc},  \textsf{zerocombine},  \textsf{flatcombine},  \textsf{cosmicray}, \textsf{xregister} and  \textsf{phot}. It performs the aperture photometry for 10 different apertures ranging from 0.7 to 4 FWHM in each images. We chose an aperture of 1.36 FWHM, which minimizes the $\Delta V$-dispersion.

\section{Results}

\subsection{CASLEO $S$-index}\label{err}

The Mount Wilson $S$-index  is defined as the ratio between the chromospheric  Ca \II\- H and K line-core emissions, integrated with a triangular profile of 1.09 \AA\- FWHM and the photospheric continuum fluxes integrated in two 20 \AA passband centred at 3891 and 4001 \AA \citep{Duncan91}. We used this expression to compute the $S$ index for the CASLEO spectrum, and we calibrated to the Mount Wilson index with Eq. (6) in \cite{Cincunegui07b}.

\begin{figure}
  \centering
  \includegraphics[width=\columnwidth]{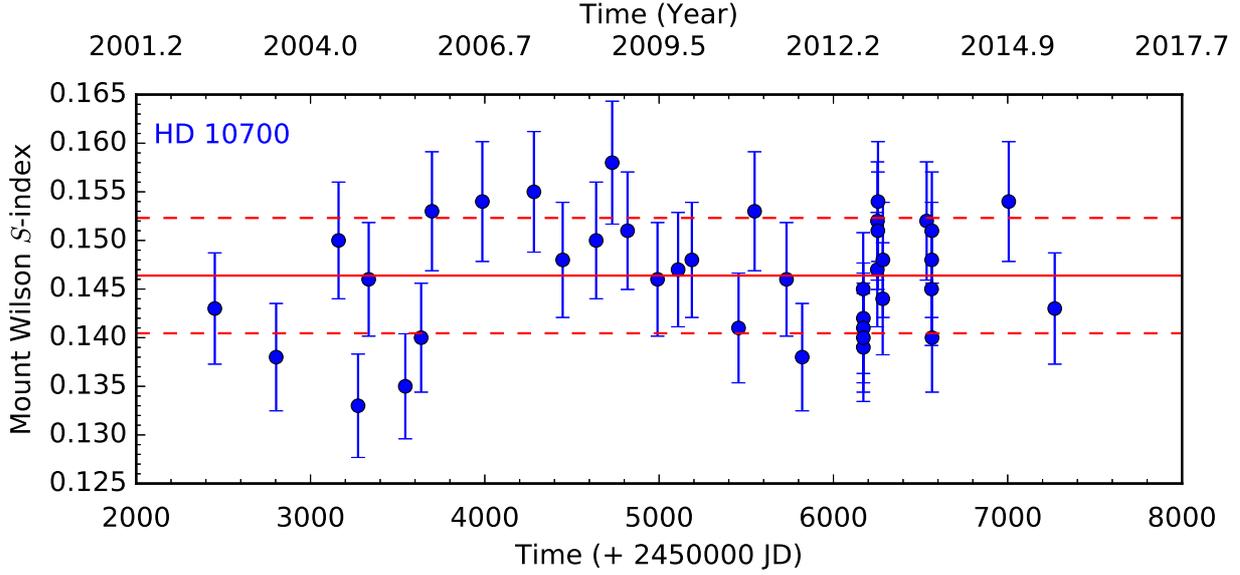}
  \caption{Mount Wilson $S$-index  obtained from CASLEO spectra for $\tau$Ceti (HD 10700) between 2002 and 2015. The solid line indicates the mean value of the series and the dotted lines the $\pm 1\sigma$ levels.  }\label{tauceti}
\end{figure}

To quantify the typical dispersion of our observations, we studied the long-term activity of the solar-type star G8.5V $\tau$ Ceti (HD 10700) classified as a \textit{flat} star by \cite{Baliunas95}. In Fig. \ref{tauceti} we plot the time series of the $S$ index of HD 10700 derived from CASLEO spectra between 2002 and 2015. We show the mean value $\langle S\rangle$ with a solid line and the $\pm 1\sigma_S$ levels with dashed lines. We consider the dispersion of this series $1\sigma_s/\langle S \rangle\sim 4\%$ as an estimation of the typical error of the activity index $S$ derived from CASLEO observations. 

\subsection{Long-term activity of AU Mic} \label{spec}

All HARPS, FEROS and UVES Ca \II\- indexes were computed following
\citeauthor{Duncan91}'s expression. We derived the Mount Wilson
$S$-index for the 31  HARPS spectra by using the calibration procedure
explained in \cite{Lovis11}, while for the 6 FEROS spectra we employed the 
\cite{Jenkins08} calibration. 
To calibrate the 3 UVES indexes, we calculate the difference between indexes  of simultaneous observations with respect to HARPS. In the case of CASLEO the calibration of the spectra is done with the \cite{Cincu04} method. In addition to this, we perform the intercalibration between $S$ indexes where the CASLEO indexes were corrected by a factor of 1.07 to re-scale Mount Wilson indexes derived from CASLEO to simultanoues measurements obtained from HARPS.

\begin{figure}
  \centering
\subfigure[\label{chfu}]{  \includegraphics[width=\columnwidth]{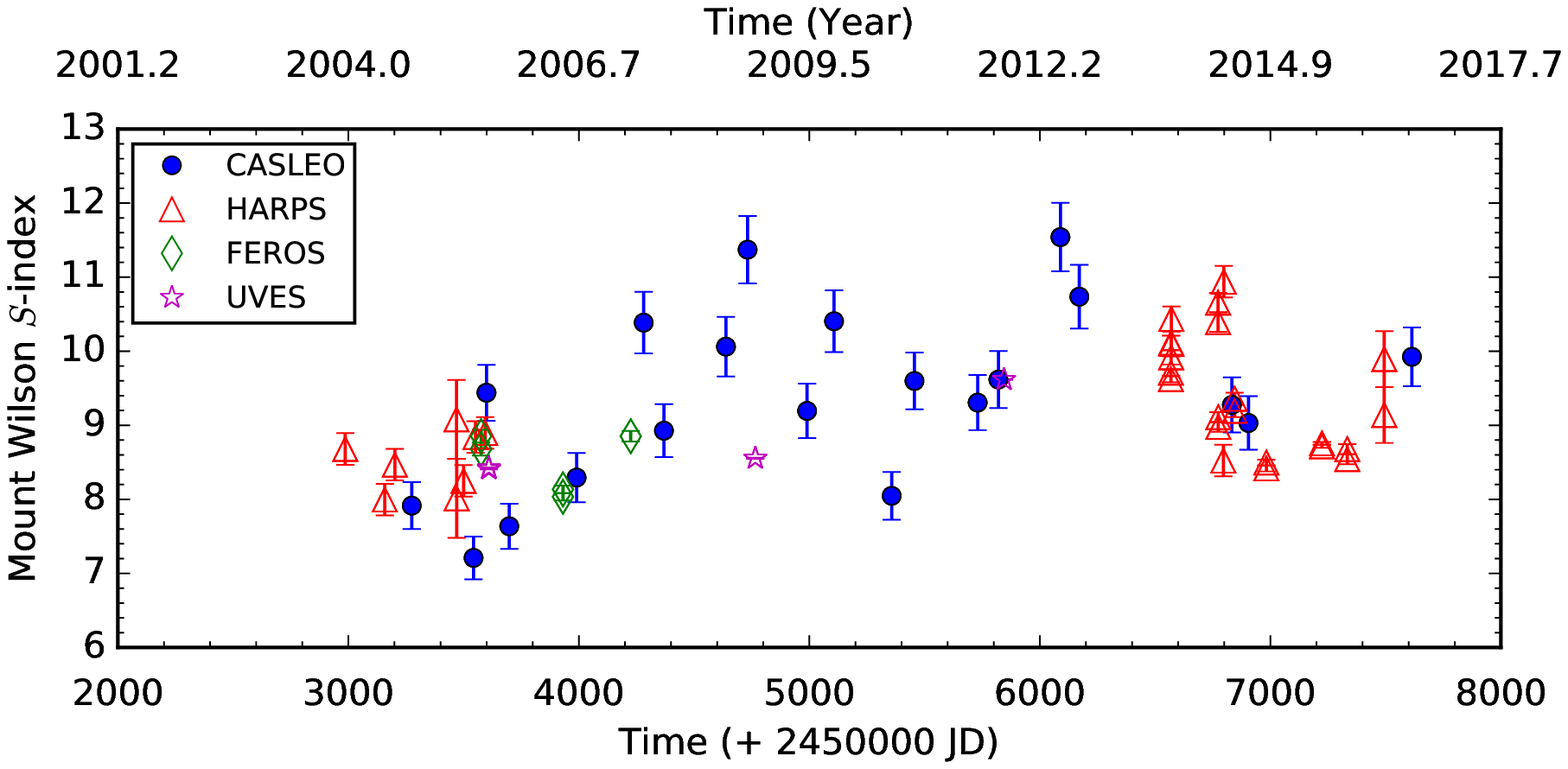}}
\subfigure[\label{perchfu}]{  \includegraphics[width=\columnwidth]{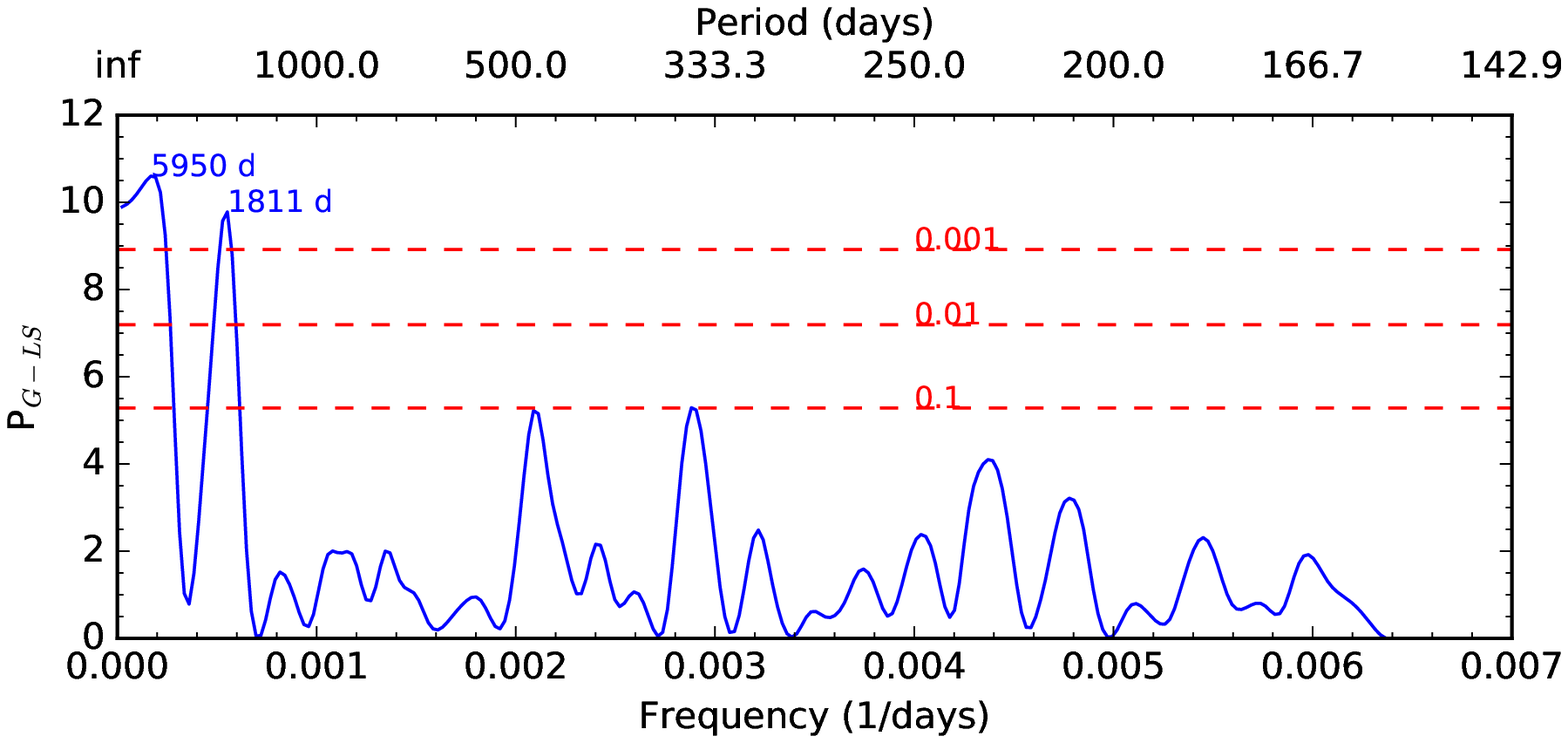}  }
  \caption{In Fig. \ref{chfu},  Mount Wilson indexes $S$ for AU Mic
    derived from CASLEO ($\bullet$), HARPS ($\bigtriangleup$), FEROS
    ($\diamond$) and UVES ($\star$) spectra. In Fig. \ref{perchfu},
    the GLS periodogram of the time series. The most significant peaks
    are indicated. The dashed lines indicated the False Alarm Probability levels. } 
\end{figure}

In Fig. \ref{chfu} we show the Mount Wilson indexes compiled for AU
Mic between 2004 and 2016 at CASLEO, combined with the Mount Wilson
indexes obtained from HARPS, FEROS and UVES spectra between
2003 and 2015.  

As we said, we assume a 4\%-error for the Mount Wilson indexes derived from CASLEO
spectra. Error bars of each HARPS, FEROS and UVES
monthly means  correspond to their standard deviation. For
those cases with only one observation in a month, we adopted the
typical RMS dispersion of other bins. 

From the whole time-series we obtained a mean Mount Wilson index
$\langle S\rangle=9.17$ with a  10\%-variability given by
$\sigma_S=0.93$. Using the \cite{Astudillo17} calibration, we obtained a
$\langle log(R'_{HK})\rangle=-3.989$, corresponding to a very active
star.

To search for stellar activity cycles, the Lomb-Scargle (LS)
periodogram \citep{Horne86} has been  extensively employed by several
studies (e.g. \citealt{Baliunas95,Metcalfe13,Flores17}). Although the
LS periodogram is a well-known algorithm to perform a frequency
analysis of  unevenly-sampled data-sets, its efficiency has been extensively
discussed in the literature and new modifications were proposed.  One
of them is the Generalized Lomb-Scargle (GLS) periodogram
\citep{Zechmeister09}, which exhibits certain  advantages in
comparison to  the classic LS periodogram. It takes into 
account a varying zero point, it does not require a bootstrap or
Monte Carlo algorithms to compute the significance of a signal,
reducing the computational cost, and it is less
susceptible to aliasing than the LS periodogram. 

In Fig.\ref{perchfu} we show the GLS periodogram for the combined spectroscopy. To
estimate the false alarm probabilities (FAP) of each peak of
the periodogram we use 

$$FAP = 1- \left[ 1 - \left( 1 - \frac{2P}{N-1} \right)^{\frac{N-3}{2}}  \right] ^M$$ 
where $N$ is the size of the data-set, $M$ is the number of independent
frequencies and $P$ is the power of the period detected in the GLS
periodogram \citep{Zechmeister09}. We normalize the periodogram
assuming Gaussian noise.  

The two most significant peaks we obtained are $P = (5950 \pm 2565)$              
days ($\sim$ 16.7 years) and $P = (1811 \pm 82)$ days ($\sim$ 5 years)
with very good FAPs of 0.02 and 0.06 \%, see Fig. \ref{perchfu}. The
error of the period detected depends on the finite frequency
resolution of the periodogram $\delta\nu$ as given by Eq.(2) in
\cite{Lamm04}, $\delta P=\frac{\delta\nu P^2}{2}$.



We also analyzed the data with other formalisms. First, we implemented the Bayesian formalism for the Generalized Lomb-Scargle periodogram (BGLS; \citealt{Mortier15}), which has the advantage that the relative probabilities of any frequency can be easily calculated and it is useful to discern the most significant peaks in the data. In this case we found a most probable period $P_{BGLS} = 1841$ days with a probability of 92 \%, and a second period of 3885 days with 47\% of probability.

In addition, we used the Phase Dispersion Minimization algorithm (PDM;
\citealt{Stellingwerf78}) provided by IRAF, which finds the frequency
that produces the smaller scatter in the phased
series. In this case we obtained two periods of 1872 and 4950 days
with a significance of $\Theta_1 = $0.31 and $\Theta_2 = $0.51 respectively. Here $\Theta$ is defined
as $s^2 / \sigma^2$  where $s^2$ is the sample variance of data
subsets, obtained by splitting the data series into
several sub samples, and $\sigma^2$ is the variance of the original
data series. For the correct period $\Theta$ should approach zero, and 
$\Theta \sim 1$ implies that the period obtained is false.

In Table \ref{tabla_periodos} we summarize the results obtained with
all methods. We see that the periods obtained by the different formalisms are
consistent with each other. We remark that each BGLS and PDM methods detect the best period without informing its uncertainty.  To verify these results with an independent data set,
we also analyzed the ASAS photometry. 

\begin{table}
\centering
\caption{Activity cycle periods detected by the different methods.}
\begin{adjustbox}{max width=\textwidth}
\begin{tabular}{ccccc}
\hline\hline\noalign{\smallskip}
Method		&	Short Period	 &	Significance &	Long Period	&	Significance \\
			&	(days)		 &	(\%)  		 &	(days)		&	(\%)       \\
\hline\noalign{\smallskip}
GLS  & 1811 $\pm$ 82   & 0.06$^*$   & 5950 $\pm$ 2565 & 0.02$^*$   \\
BGLS & 1841	           & 92$^{**}$  & 3885	         & 47$^{**}$        \\
PDM  & 1872            & 31$^{***}$ & 4950            & 51$^{***}$   \\
ASAS & 1929 $\pm$ 283  & 6          &				 & \\

\hline 
\end{tabular}
\end{adjustbox}
\centering

$^*$ False Alarm Probability (FAP). $^{**}$ Relative probability. $^{***}$ Significance $\Theta$.

\label{tabla_periodos}
\end{table}

\begin{figure}
  \centering
\subfigure[\label{asas}]{  \includegraphics[width=\columnwidth]{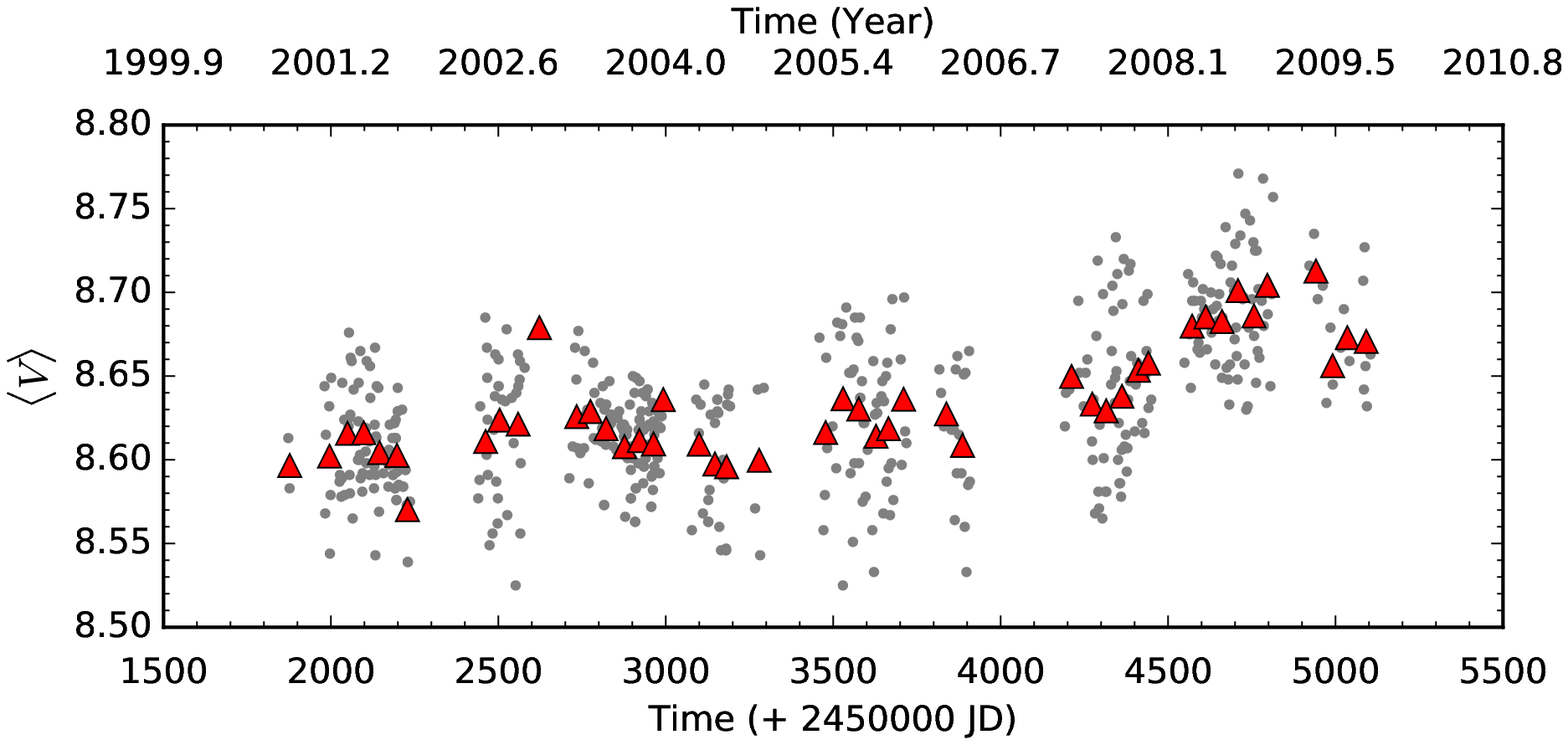}}
 \subfigure[\label{perasas}]{  \includegraphics[width=\columnwidth]{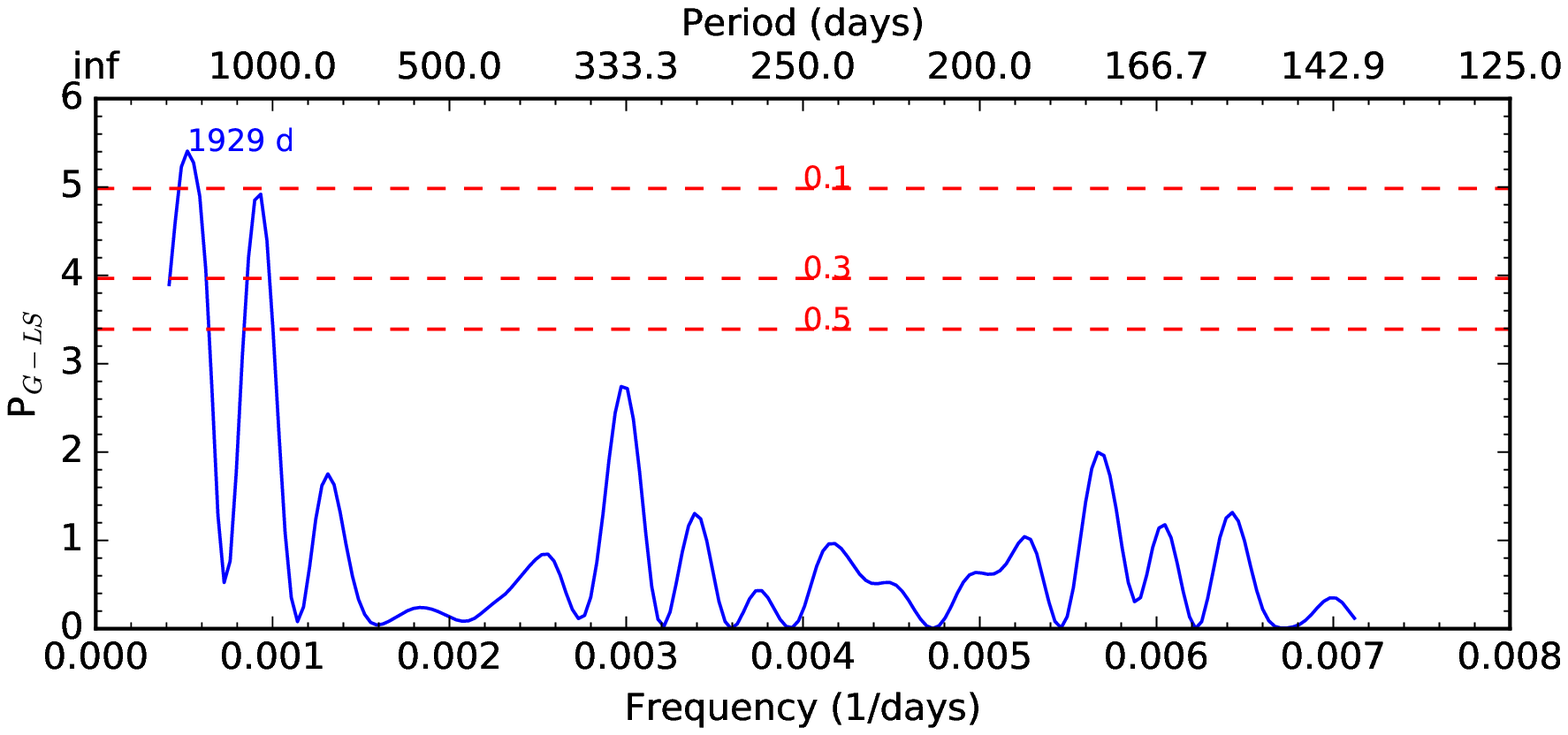}}
  \caption{In Fig. \ref{asas}, the $V$ magnitude for AU Mic given by
    the ASAS public database (gray circles) and the corresponding
    seasonal means (red triangles). In Fig. \ref{perasas}, the GLS
    periodogram of the seasonal mean. Both, the most significant peak and the FAP-levels (red dashed lines) are indicated.} 
\end{figure}

In Fig. \ref{asas}, we show the $V$ magnitude of AU Mic obtained from
the high-quality ASAS data-set between 2000 and 2009. The final time
series consists of 507 points, with typical errors of  around  30
mmag.  The mean magnitude of the whole data set is $\langle V \rangle
= 8.63 \pm 0.04$. To eliminate possible short-term variations due to flare events, we binned the data by grouping every 45 days. 
The red triangles represent the mean of the observations  taken
over 45 days, weighted by the error reported in the ASAS database. We computed  the error of each
mean $V$ magnitude  as  the  square  root  of  the  variance-weighted
mean (see \citealt{Frodesen79}, Eq. 9.12). 

In Fig. \ref{perasas} we show the periodogram for the ASAS data
set. For the binned series we obtained a peak at $P_{ASAS} = 1929 \pm
283$ days with a FAP = 6\%. 
As in the spectroscopic study, we also studied the smoothed data set from ASAS with the BGLS and PDM formalisms. We obtained similar periods although their significances were very low due to the high dispersion of the original data series. 

\textbf{ \cite{SuarezMascareno16} detected a period of 7.6-yr in a similar ASAS dataset. 
Thus in this work, we evaluated the original data (without binning) and the grouped data (binning the time series into different groups). We analyzed all the time-series with the GLS periodogram, but  we did not find the period of 7.6 years reported by Su\'arez Mascare\~no and collaborators. In all the cases (original and smoothed data) we found a period close to the 1929 days as we reported before.
}

\subsection{Rotation period}

The rotation period of AU Mic was determined by several authors,
with little discrepancies. First, from the ASAS database
\cite{Pojmanski02} determined a rotation period of 4.822 days. 
\cite{Hebb07} found a significant period of 4.847 days using
photometric observations obtained with the CTIO-1m telescope,
employing four different optical filters. Finally,
\cite{Kiraga12} analyzed a longer ASAS series and  determined a
rotation period of $P_{rot}=4.842$ days. 

\begin{figure}
\centering
\includegraphics[width=\columnwidth]{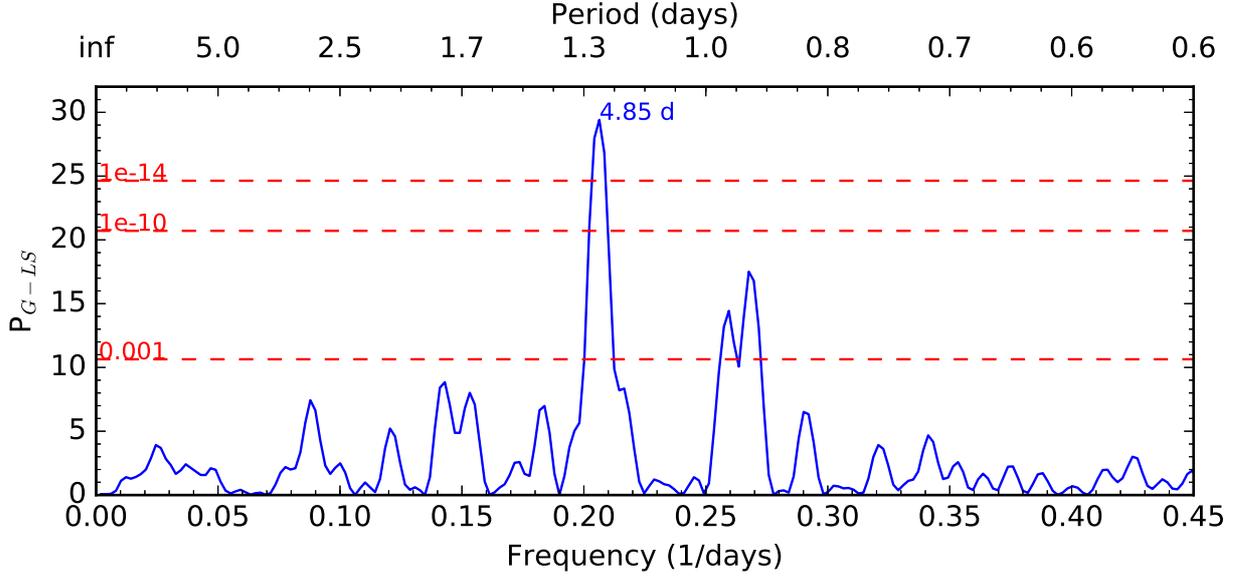}
\caption{GLS Periodogram of the binned $\Delta$V obtained with MATE.}\label{per_mate}
\end{figure}

As explained in section \S\ref{sec.obs}, we also performed differential
photometry of AU Mic. Analyzing  the time series of 1902 points obtained as $\Delta V$  between the AU Mic and its comparing star with a GLS
periodogram, we found a rotation period of $P = (4.843 \pm
0.004)$ days with a very low FAP.
However, this periodogram presents other significant peaks, probably
of instrumental origin or due to aliasing. For this reason, we binned
the data corresponding to each night in three points, corresponding
to the beginning, the middle and the end of the night. For the grouped data we
obtained a period of $P_{group} = (4.85 \pm 0.02)$ days with a FAP =
$5\times10^{-21}$. This periodogram, for the binned  data,
is plotted in Fig. \ref{per_mate}. 

We also studied the binned data set with the BGLS and PDM formalisms, 
described in Section $\S$ \ref{spec}. We obtained similar results with
peaks at P = 4.849 days and P = 4.89 days, respectively.

\begin{figure}
  \centering
  \includegraphics[width=\columnwidth]{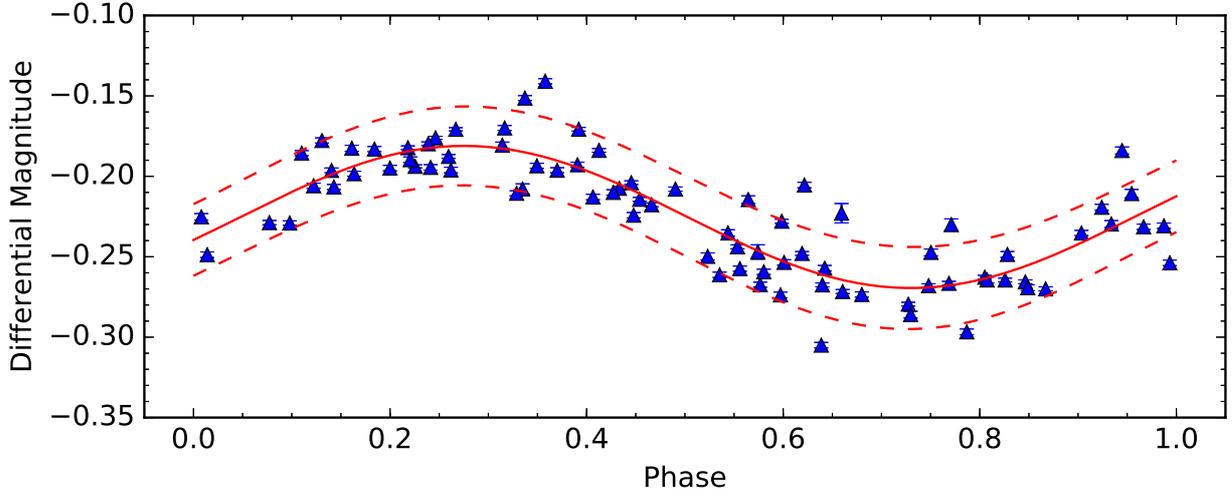}
  \caption{$\Delta$V vs. rotational phase. The dashed lines represent $\pm 3\sigma$ deviations.}
  \label{phase_mate}
\end{figure}

In Fig. \ref{phase_mate} we show the differential magnitude of the
smoothed data  phased with the rotation period  of 4.85 days, the most
significant peak in Fig. \ref{per_mate}. The red solid line represents
the least-square fit with a harmonic function.  

We did not detect signs of flare activity during our observations. In
the future, we plan to obtain  observations in the V and R filters and
higher cadence to explore the flare activity of AU Mic in detail.

\section{Discussion}

Long-term activity has been studied for only a few dM stars, where activity cycles with periods between 1 to 8 years were detected. To increase this statistics, we studied in detail the long-term activity of the active fast-rotator  dM star AU Microscopii (GJ 803-M1V). Furthermore, since most early-M stars are inactive \citep{Reiners12}, this star is an uncommon target to analyze stellar activity. Its high activity is very likely related to its  youth, as \cite{Reiners12}  also observed in other six early-M  stars.

In this work, we built for the first time a long time series of AU Mic
Mount Wilson indexes $S$ measured from CASLEO, UVES, FEROS and HARPS
spectra obtained between 2004 and 2016.
We analyzed this data with the Generalized Lomb Scargle (GLS)
periodogram and we detected two significant activity cycles of
$P \sim 1811$ days and $P \sim 5950$ days ($\sim 5$ and $\sim 16.7
$ years). 
We analyzed the same series with two other techniques and we 
obtained similar results. See Table \ref{tabla_periodos} for a summary
of the results.

We also studied ASAS photometry of this star obtained between 2000 and 2009.
With the GLS we detected an activity cycle of (1929 $\pm$ 283) days. 
Since both data-sets are completely independent, the FAP of the
combined detection of the shorter cycle is 3.6$\times 10^{-5}$.

A different case is the one of the longer period. Although the
detection in the GLS is significant, with a $FAP=0.02\%$, the error of
the period itself is quite large, and the detection of this period with the
two other methods is less significant. This is probably due to the
fact that this period is longer than the time-series. This period is
not detected in the ASAS observations, which is not surprising since
these observations span only 9 years.

To study whether this peak is an artifact caused by the duration of
the data-set, we follow \cite{Buccino09}. We used the sunspot number
$S_N$ between 1900 and 2000 obtained from the National Geophysical Data
Center\footnote{https://www.ngdc.noaa.gov/stp/solar/ssndata.html}.
We rescaled in time this series to the shorter period $P=1811$
days. We also rescaled the $S_N$ to obtain a time series of the
same mean value and we added gaussian noise to get the same
standard deviation of our data. In this way we built an artificial
signal representing a cycle of the same characteristics of the one
observed in AU Mic.

We then ``observed'' the rescaled solar data selecting data-points
with the same phase intervals than in the $S$-index time series for AU
Mic, and we computed the GLS periodogram. We repeated this
``observation'' 1000 times with random starting dates. For each
periodogram, we considered the most significant period.  As expected,
97.5 \% of the periods detected were between 4.2 and 5.7 years.  Only
in 1.7 \% of the cases this period lied between 13 and 18 years with
FAP$<$0.02.  In other words, the FAP for the longer period in this
data-set is only of 1.7 \%. Therefore, the longer period we detected
in Fig. \ref{perchfu} is likely related to stellar
activity. However, as this period presents a large error, we
  will continue observing this star in the HK$\alpha$ project, to
  obtain a longer time-series that could confirm this detection.

In addition, we also obtained our own photometry in the V-band during
several months for this star. From these observations, we obtained a
rotation period of 4.85 days with a 0.4\% accuracy. This
result agrees with rotation periods reported in the literature,
obtained from other independent data-sets
\citep{Pojmanski02,Hebb07,Kiraga12}.

It is well known that stellar rotation plays an important role on the
dynamo action in FGK stars and thus on their stellar activity
(e.g. \citealt{Noyes84}, \citealt{Mamajek08}).  
Recently, \cite{Astudillo17} provided the  activity index $R'_{HK}$
for a set of M0-M6 stars and evaluated the relation between $R'_{HK}$
and the rotation period. They obtained that the $R'_{HK}$ index shows
a saturated correlation with rotation for stars with $P_{rot}< $10
days. Above that period, the  $R'_{HK}$- $P_{rot}$ seems to be linear,
where activity decreases with slower rotation, what is found for FGK
stars. This saturation regime in M fast-rotators is already known  for
other activity proxies (eg. $L_\alpha/L_{bol}$ in
\citealt{Delfosse98}, $L_{Ca}/L_{bol}$ in \citealt{Browning10}, $B_f$
in \citealt{Reiners09}). While the whole pattern  observed in the
$R'_{HK}$- $P_{rot}$ relation    was also reported  in the X-ray
coronal emission $L_x/L_{bol}$ by \cite{Wright16}. 

From our Mount Wilson series we obtained an activity index  $\langle log(R'_{HK})\rangle=-3.989 $. Given a rotation period of $P_{rot}$= 4.85 days, this $\langle log(R'_{HK})\rangle$ value place AU Mic  in the saturation regime, as is also observed for the $L_X/L_{bol}$ proxies \citep{Wright11}. It reflects the fact that the rotation plays a key role in the magnetic activity on AU Mic similar to other young FGK stars, thus its dynamo mechanism should be similar to a solar-type dynamo. Moreover, at an age of about 25\,Myr, stars of M1V spectral type like AU Mic are not fully convective any more, with an external convective envelope whose mass ranges from 40\% \citep{Baraffe15} to 54\% \citep{Feiden16} of the total mass, depending on the adopted evolutionary model. Therefore, we can expect to have in AU Mic mechanisms of magnetic fields generation similar to those that operate in the more massive solar-type stars. Therefore the $\sim$ 5-year chromospheric activity cycle detected for AU Mic is consistent with an $\alpha \Omega$ dynamo.  \cite{Saar94} determined a magnetic field strength for AU Mic $B_f=2.3$ KG. \cite{Malo14} suggest that this magnetic field strength is consistent with the one derived from an $\alpha-\Omega$ dynamo model for AU Mic given its age approximately. Considering a Rossby Number $R_0\sim$0.15 \textbf{\citep{Wright16}}, this star seems to be near the limit of the saturation regime of the $B_f$  proxy. Therefore, we bring new evidence that there is not a tight correlation between emission and magnetic fields as observed in other active stars \citep{Reiners09}. 

On the one hand, the determination of the activity cycle and rotation periods
of AU Mic, allows us to analyze its $P_{cyc}-P_{rot}$
relation in the stellar context. For those FGK stars latter than F5 with
well-determined rotation and cycle periods, \cite{Bohm07} examined this
relation. She concluded that the values of $P_{cyc}$ are
distributed in two branches in the $P_{cyc}-P_{rot}$
diagram, indicating that probably different kinds of dynamos are
operating in each different sequence.

Expanding this sample, \cite{Saar11}   analyzed the dependency of the cycle frequency ($\omega_{cyc}$)  on the rotation frequency ($\Omega$) for a set of single active F5-M4 stars. They identified three branches on the $\omega_{cyc}-\Omega$ diagram for stars with rotations bellow  10 $\Omega_\odot$.  We included the rotation and cyclic periods  detected in the present work on the diagrams reported in \cite{Bohm07}  and \cite{Saar11}. In both cases we observe that the $\sim$5-year activity cycle   belongs to the \textit{active} branch, which means that particular our results for AU Mic are consistent with the $\omega_{cyc}-\Omega$ distribution for several stars.


Following Saar's diagram, if AU Mic also has a shorter cycle, as reported in other M stars \citep{Buccino14}, it would belong to the \textit{inactive}
branch, therefore, should have a period between 1.14 and 1.57 years.
In our GLS periodogram of the Mount Wilson indexes a poor significant
peak (FAP=1\%) at 1.36 years is present. Due to the low significance
of this detection, we think that a 2-year observing campaign with
weekly observations  is needed to really detect this second shorter
cycle of lower amplitude. 



On the other hand, since AU Mic is surrounded by an edge-on circumstellar debris disk \citep{Kalas04,Liu04,Krist05}, our results could be relevant on characterizing this star as a planetary host.  Until 2015 only three debris disks were detected surrounding M stars: AU Mic, GJ 581 \citep{Lestrade12} and the M4 star Fomalhaut C \citep{Kennedy14}. Recently, \cite{Choquet16} reported the detection of two disks imaged for the first time in scattered light around the M stars: TWA 7 (M3.2, 34.5 pc, 10 Myr) and TWA 25 (M0.5, 54 pc, 13 Myr). The presence of a debris disk indicates that its stellar system achieved the formation of planetary-like bodies during its previous protoplanetary disk phase. In the case of GJ 581 five planets were detected around this star (\citealt{Bonfils05,Udry07,Mayor09}).

The AU Mic's youth and its proximity to Earth ($\sim$ 9.9 pc), are favourable conditions for the detection of low mass planets. \cite{Boccaletti15} reported five large-scale features with a fast outward motion of the AU Mic disk  that can be interpreted as signposts of significant planetary formation activity. Therefore, an accurate determination of the rotation period and the activity cycles of AU Mic, reported for the first time in this present work, should be useful for planets searches on this star since signals related to stellar activity could bring false detections in different time-scales \citep{Dumusque11,Diaz16}.

\bibliographystyle{mnras}
\small
\bibliography{biblio}

\end{document}